\documentclass[journal=jcisd8,manuscript=article,layout=traditional]{achemso}

\usepackage[version=3]{mhchem} %
\usepackage{amsmath,amssymb,bm}
\usepackage{multirow}
\usepackage{siunitx}
\usepackage{graphicx}
\usepackage{subcaption}
\usepackage[justification=centering]{caption}
\usepackage{longtable}
\usepackage{xcolor}

\newcommand{\tss}[1]{\textsuperscript{#1}}

\author{Mostafa Karimi}
\affiliation[Texas A\&M University]
{Department of Electrical and Computer Engineering, Texas A\&M University, College Station, TX 77843, United States}
\alsoaffiliation[Texas A\&M University]
{TEES-AgriLife Center for Bioinformatics and Genomic Systems Engineering, Texas A\&M University, College Station, TX 77840, United States}
\altaffiliation{Co-first authors}
\author{Di Wu}
\affiliation[Texas A\&M University]
{Department of Electrical and Computer Engineering, Texas A\&M University, College Station, TX 77843, United States}
\altaffiliation{Co-first authors}
\author{Zhangyang Wang}
\affiliation{Department of Computer Science and Engineering, Texas A\&M University, College Station, TX 77843, United States}
\author{Yang Shen}
\email{yshen@tamu.edu}
\affiliation[Texas A\&M University]
{Department of Electrical and Computer Engineering, Texas A\&M University, College Station, TX 77843, United States}
\alsoaffiliation[Texas A\&M University]
{TEES-AgriLife Center for Bioinformatics and Genomic Systems Engineering, Texas A\&M University, College Station, TX 77840, United States}

\title[Explainable Prediction of Compound-Protein Interactions]{Explainable Deep Relational Networks for Predicting Compound-Protein Affinities and Contacts}

\begin{document}

\begin{abstract}

Predicting compound-protein affinity is critical for accelerating drug discovery. Recent progress made by machine learning focuses on accuracy but leaves much to be desired for interpretability.  Through molecular contacts underlying affinities, our large-scale interpretability assessment finds commonly-used attention mechanisms inadequate. We thus formulate a hierarchical multi-objective learning problem whose predicted contacts form the basis for predicted affinities.  We further design a physics-inspired deep relational network, DeepRelations, with intrinsically explainable architecture.  Specifically, various atomic-level contacts or ``relations'' lead to molecular-level affinity prediction.  And the embedded attentions are regularized with predicted structural contexts and supervised with partially available training contacts.  DeepRelations shows superior interpretability to the state-of-the-art: without compromising affinity prediction, it boosts the AUPRC of contact prediction 9.5, 16.9, 19.3 and 5.7-fold for the test, compound-unique, protein-unique, and both-unique sets, respectively. Our study represents the first dedicated model development and systematic model assessment for interpretable machine learning of compound-protein affinity.

\end{abstract}

\section{Introduction}

Current drug-target interactions are predominantly represented by the interactions between small-molecule compounds as drugs and proteins as targets \cite{santos2017comprehensive}.  The enormous chemical space to screen compounds is estimated to contain $10^{60}$ drug-like compounds \cite{chemspace96}.  And these compounds act in biological systems of millions or more protein species or ``proteoforms'' (considering genetic mutations, alternative splicing, and post-translation modifications of proteins)  \cite{proteoforms1,proteoforms2}.  Facing such a combinatorial explosion of compound-protein pairs, drug discovery calls for efficient characterization of compound efficacy and toxicity, and computational  prediction of compound-protein interactions (CPI) addresses the need.  

Recently computational CPI prediction has made major progress beyond predicting whether compounds and proteins interact.  Indeed, thanks to increasingly abundant molecular data and advanced deep-learning techniques, compound-protein affinity prediction is reaching unprecedented accuracy, with inputs of compound-protein structures \cite{wallach2015atomnet,gomes2017atomic,jimenez2018k}, compound identities (such as SMILES and graphs) and protein structures (see a relevant problem of binding classification\cite{torng2019graph,lim2019predicting}), or even just compound-protein identities \cite{karimi2018deepaffinity,ozturk2018deepdta,PADME}.  As previously summarized \cite{karimi2018deepaffinity}, structure-based affinity-prediction methods are limited in applicability due to the often-unavailable structures of compound-protein pairs or even proteins alone, whereas structure-free methods, being broadly applicable, could be limited in interpretability.

Interpretability remains a major gap between the capability of current compound-affinity predictors and the demand of rational drug discovery.  The central question about interpretability is whether and how methods (including machine learning models) could explain \textit{why} they make certain predictions (affinity level for any compound-protein pair in our context).  This important topic is rarely addressed for affinity prediction.  DeepAffinity \cite{karimi2018deepaffinity} has embedded joint attentions over compound-protein component pairs and uses such joint attentions to assess origins of affinities (binding sites) or specificities.  Additionally, attention mechanisms have been used for predictions of CPI \cite{gao2018interpretable}, chemical stability \cite{li2019deepchemstable} and protein secondary structures \cite{uddin2019saint}.  Assessment of interpretability for all these studies was either lacking or limited to few case studies.  We note a recent work proposing \textit{post-hoc} attribution-based test to determine whether a model learns binding mechanisms \cite{mccloskey2019attr}.  

We raise reasonable concerns on how much attention mechanisms can reproduce natural  contacts in compound-protein interactions. Attention mechanisms were originally developed to boost the performance of seq2seq models for neural machine translations \cite{bahdanau2014neural}. And they have gained popularity for interpreting deep learning models in visual question answering \cite{lu2016hierarchical}, natural language processing \cite{xu2015show} and healthcare \cite{choi2016retain}. However, they were also found to work differently from human attentions in visual question answering \cite{das2017human}.  

Representing the first effort dedicated to interpretability of compound-protein affinity predictors, our study is focused on how to define, assess, and enhance such  interpretability as follows.  

\textit{How to define interpretability for affinity prediction.}  Interpretable machine learning is increasingly becoming a necessity \cite{doshivelez2017rigorous} for fields beyond drug discovery.  Unlike interpretability in a generic case \cite{doshivelez2017rigorous}, what interpretability actually means and how it should be evaluated is much less ambiguous for compound-protein affinity prediction.  So that explanations conform with scientific knowledge, human understanding, and drug-discovery needs, we define interpretability of affinity prediction as the ability to explain predicted affinity through underlying  atomic interactions (or contacts).  Specifically, atomic contacts of various types are known to constitute the physical basis of intermolecular interactions \cite{Dill2012Book}, modeled in force fields to estimate interaction energies \cite{gilson2007calculation}, needed to explain mechanisms of actions for drugs \cite{brzozowski1997molecular,congreve2005keynote}, and relied upon to guide structure-activity research in drug discovery \cite{wlodawer1998inhibitors,brik2003hiv}.  We emphasize that simultaneous prediction of affinity and contacts does not necessarily make the affinity predictors  intrinsically interpretable unless predicted contacts form the basis for predicted affinities.

\textit{How to assess interpretability for affinity prediction.}  Once interpretability of affinity predictors is defined through atomic contacts, it can be readily assessed against ground truth known in compound-protein structures, which overcomes the barrier for interpretable machine learning without ground truth \cite{yang2019evaluating}. In our study, we  have curated a dataset of compound-protein pairs, all of which are labeled with $K_d$ values and some of which with contact details; and we have split them into training, test, compound-unique, protein-unique, and both-unique (or double-unique) sets. We measure the accuracy of contact prediction over various sets using area under the precision-recall curve (AUPRC) which is suitable for binary classification (contacts/non-contacts) with imbalanced classes (far less contacts than non-contacts). We have performed large-scale assessments of attention mechanisms in various molecular data representations (protein amino-acid sequences and structure-property annotated sequences\cite{karimi2018deepaffinity} as well as compound SMILES and graphs) and corresponding neural network architectures (convolutional and recurrent neural networks [CNN and RNN] as well as graph convolutional and isomorphism networks [GCN and GIN]).  And we have found that current attention mechanisms inadequate for interpretable affinity prediction, as their AUPRCs were merely 50\% more than chance.  

\textit{How to enhance interpretability for affinity prediction.}  We have made three main contributions to enhance interpretability.  

The first contribution, found to be the most impactful, is to design intrinsic  explainability into the architecture of a deep ``relational'' network. Inspired by physics, we explicitly model and learn various types of atomic interactions (or ``relations'') through deep neural networks and embed attentions at the levels of residue-atom pairs and relation types.  This was motivated by relational neural networks first introduced to learn to reason in computer vision \cite{santoro2017simple,lu2016visual} and subsequent interaction networks to learn the relations and interactions of complex objects and their dynamics \cite{battaglia2016interaction,hoshen2017vain}.  Moreover, we combine such deep relational modules in hierarchy to progressively focus attention from putative protein surfaces, binding-site $k$-mers and residues, to putative residue-atom binding pairs.  

The second contribution is to incorporate physical constraints into data representations, model architectures, and model training.  (1) To respect the sequence nature of protein inputs and to overcome the computational bottlenecks of RNNs, inspired by protein folding principles, we represent protein sequences as hierarchical $k$-mers and model them with hierarchical attention networks (HANs).  (2) To respect the structural contexts of proteins, we predict from protein sequences solvent exposure over residues and contact maps over residue pairs; and we introduce novel structure-aware regularizations for structured sparsity of model attentions.  

The third contribution is to supervise attentions with partially available contact data and train models accordingly.  For interpretable and accurate affinity prediction, we have formulated a hierarchical multi-objective optimization problem where contact predictions form the basis for affinity prediction. We utilize contact data available to a minority (around 7.5\%) of training compound-protein pairs and design hierarchical training strategies accordingly.   

The rest of the paper is organized as follows.  The aforementioned contributions in defining, measuring, and enhancing intepretable affinity prediction will be detailed in Methods.  In Results, compared to state-of-the-art first, the resulting framework of DeepRelations is found to drastically boost interpretability robustly over default test, protein-unique, compound-unique, and double-unique sets, without sacrificing accuracy. Ablation studies then reveal the most contributing methodological contribution --- the intrinsically explainable model architecture of our deep ``relational'' networks.  Case studies provide further insights into the pattern of interpreted contacts.

\section{Methods}

Toward genome-wide prediction of compound-protein interactions (CPI), we assume that proteins are only available in 1D amino-acid sequences, whereas compounds are available in 1D SMILES or 2D chemical graphs.  We start the section with the curation of a dataset of compound-protein pairs with known $pK_d$ values, a subset of which is of known intermolecular contacts. We will introduce the state-of-the-art and our newly-adopted neural networks to predict from such molecular data.  These neural networks will be first adopted in our previous framework of DeepAffinity \cite{karimi2018deepaffinity} (supervised learning with joint attention) so that the interpretability of attention mechanisms can be systematically assessed in CPI prediction.  We will then describe our physics-inspired, intrinsically explainable architecture of deep relational networks where aforementioned neural networks are used as basis models. With carefully designed regularization terms, we will explain multi-stage deep relational networks that increasingly focus attention on putative  binding-site $k$-mers, binding-site residues, and residue-atom interactions, for the prediction and interpretation of compound-protein affinity.  We will also explain how the resulting model can be trained using  compound-protein pairs with affinity values but not necessarily with atomic interaction details.  

\subsection{Curation of a CPI Relational Benchmark Set}
We have previously curated affinity-labeled compound-protein pairs \cite{karimi2018deepaffinity} based on BindingDB \cite{liu2006bindingdb}. The compound data were in the format of canonical SMILES as provided in PubChem \cite{kim2018pubchem} and the protein data are in the format of FASTA sequences.  In this study, we used those data with amino-acid sequences no more than 300 from the p$K_d$-labeled set \cite{karimi2018deepaffinity}, which corresponds to 1,926 compound-protein pairs. We also converted SMILES to graphs with RDKit~\cite{rdkit}.    

The p$K_d$-labeled data only shows the affinity strength between proteins and compounds, but it lacks the details on where and how the pairs interact. We have thus curated a subset of the p$K_d$-labeled data with atomic-level intermolecular contacts (or ``relations") derived from compound-protein co-crystal structures in PDB \cite{berman2000protein}, as ground truth for the interpretablity of affinity prediction.  Specifically, we cross-referenced aforementioned compound-protein pairs in PDBsum\cite{laskowski2018pdbsum} and used its LigPlot service to collect high-resolution atomic contacts or relations.  These relations are given in the form of contact types (hydrogen bond or hydrophobic contact), atomic pairs, and atomic distances.  

The resulting dataset of 1,926 p$K_d$-labeled compound-protein pairs (including 144 pairs with atomic-contact data) corresponds to 137 proteins and 1,376 compounds.  We randomly split them into four folds where fold 1 do not overlap with fold 2 in compounds, do not do so with fold 3 in proteins, and do not do so with fold 4 in either compounds or proteins.  Folds 2, 3, and 4 are referred to as compound-unique, protein-unique, and double unique sets for generalization tests; and they contain 201(19), 191(14), and 192(10) compound pairs (including those with contact details in the parentheses).  Fold 1 was randomly split into training (70\%) and test (30\%) sets where 10\% of the training set was set aside as the validation or development set.  The training (including validation) and test sets contain 974(74) and 368(27) compound-protein pairs (with contact details).  The split of the whole dataset is illustrated in Figure~\ref{fig:dataset} below.

\begin{figure}[!htb]
    \centering
    \includegraphics[width=.7\columnwidth]{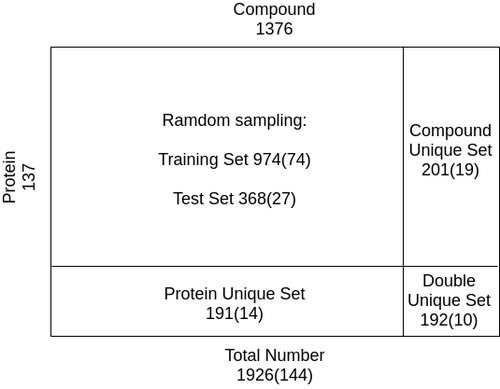}
    \caption{The complete data set consists of training, test, compound-unique, protein-unique, and double unique sets with compound-protein counts provided (including those with contact details in parentheses).}
    \label{fig:dataset}
\end{figure}

Although monomer structures of proteins are often unavailable, their structural features can be predicted from protein sequences alone with reasonable accuracy.  We have predicted the secondary structure and solvent accessibility of each residue using the latest SCRATCH \cite{cheng2005scratch,magnan2014sspro} and contact maps for residue pairs using RaptorX-contact~\cite{wang2017accurate}.  These data provide additional structural information to regularize our machine learning models.  If protein structures are available, actual rather than predicted such data can be used instead.

\subsection{Data Representation and Corresponding Basis Neural Networks}

\subsubsection{Baseline: CNN and RNN for 1D protein and compound sequences.}

When molecular data are given in 1D sequences, these inputs are often processed by convolutional neural networks (CNN) \cite{ozturk2018deepdta,lee2019deepconv} and by recurrent neural networks (RNN) that are more suitable for  sequence data with long-term interactions \cite{karimi2018deepaffinity}.  

Challenges remain in RNN for compound strings or protein sequences. For compounds  in  SMILES  strings,  the  descriptive  power  of  such  strings  can  be limited. In this study, we overcome the challenge by representing compounds
in chemical formulae (2D graphs) and using two types of graph neural networks (GNN). For proteins in amino-acid sequences, the often-large lengths demand deep RNNs that are hard to be trained effectively (gradient vanishing or exploding and non-parallel training)  \cite{trinh2018learning}.  We previously overcame the second challenge by predicting structure properties from amino-acid sequences and representing proteins as a much shorter structure property sequences where each 4-letter tuple corresponds to a secondary structure \cite{karimi2018deepaffinity}.  This treatment however limits the resolution of interpretability to be at the level of protein secondary structures (multiple neighboring residues) rather than individual residues. In this study, we overcome the second challenge while achieving residue-level interpretability by using biologically-motivated hierarchical RNN (HRNN).  

\subsubsection{Proposed: GCN and GIN for 2D compound graphs.}
Compared to 1D SMILES strings, chemical formulae (2D graphs) of compounds have more descriptive power and are increasingly used as inputs to predictive models  \cite{gao2018interpretable,PADME,karimi2018deepaffinity,li2019deepchemstable,KekuleScope2019}.  In this study, compounds are represented as 2D graphs in which vertices are  atoms and edges are covalent bonds between atoms. Suppose that $n$ is the maximum number of atoms in our compound set (compounds with smaller number of atoms are padded to reach size $n$). Let's consider a graph $G = (\mathcal{V},\mathcal{X},\mathcal{E},\mathcal{A})$,  where $\mathcal{V}=\{v_j\}_{j=1}^{n}$ is the set of $n$ vertices (each with $d_g$ features), $\mathcal{X} \in R^{n\times d_g}$ that of vertex features,  $\mathcal{E}$ that of edges, 
and $\mathcal{A} \in \{0,1\}^{n\times n}$ is unweighted symmetric adjacency matrix. Let $\hat{\mathcal{A}} = \mathcal{A}+\mathcal{I}$ and $\hat{\mathcal{D}}$ be the degree matrix (the diagonals of $\hat{\mathcal{A}}$).  

We used Graph Convolutional Network (GCN) \cite{kipf2016semi} and Graph Isomorphism Network (GIN)
 \cite{xu2018powerful} which are the state of art for graph embedding and inference. GCN consists of multiple layers and at layer $l$ the model can be written as:
\begin{equation}
    \mathcal{H}^{(l)} = \mathrm{ReLU} (\hat{\mathcal{D}}^{-\frac{1}{2}}\hat{\mathcal{A}}\hat{\mathcal{D}}^{-\frac{1}{2}}\mathcal{H}^{(l-1)}\Theta^{(l)}), 
\end{equation}
where ${\mathcal{H}^{(l)}} \in R^{n \times d_g^{(l)}}$ is the output, $\Theta^{(l)} \in R^{d_g^{(l-1)}\times d_g^{(l)}}$ the trainable parameters, and $d_g^{(l)}$ the number of features, all at layer $l$. Initial conditions (when $l=0$) are $\mathcal{H}^{(0)}=\mathcal{X}$ and $d_g^{(0)}=d_g$. 

GIN is the most powerful graph neural network in theory: its  discriminative or representational power is equal to that of the Weisfeiler-Lehman graph isomorphism test \cite{weisfeiler1968reduction}. Similar to GCN, GIN consists of multiple layers and at layer $l$ the model can be written as a multi-layer perceptron (MLP):  
\begin{equation}
    \mathcal{H}^{(l)} = \mathrm{MLP}^{(l)} (\bar{\mathcal{A}}^{(l)} \,\, \mathcal{H}^{(l-1)}),
\end{equation}
where $\bar{\mathcal{A}}^{(l)} = \mathcal{A}+ \epsilon^{(l)} \mathcal{I}$, $\epsilon^{(l)}$ can be either a trainable parameter or a fixed hyper-parameter. Each GIN layer has several nonlinear layers compared to GCN layer with just a ReLU per layer, which might improve predictions but suffer in interpretability. 

The final representation for a compound is $\mathcal{Y} = \mathcal{H}^{(L)}$ if GCN or GIN has $L$ layers.  In this study, vertex features are as in \cite{li2019deepchemstable}, with few additional features detailed later for physics-inspired relational modules.  

\subsubsection{Proposed: HRNN for 1D protein sequences.}

We aim to keep the use of RNN that respects the sequence nature of protein data and mitigate the difficulty of training RNN for long sequences. To that end, inspired by the hierarchy of protein structures, we model protein sequences using hierarchical attention networks (HANs).  Specifically, during protein folding, sequence segments may fold separately into secondary structures and the secondary structures can then collectively pack into a tertiary structure needed for protein functions. We exploit such hierarchical nature by representing a protein sequence of length easily in thousands as tens or hundreds of $k$-mers (consecutive sequence segments) of length $k$ ($k=15$ in this study). Accordingly we process the hierarchical data with hierarchical attention networks (HANs) \cite{yang2016hierarchical} which have been proposed for natural language processing.  We also refer to it as hierarchical RNN (HRNN).   

Given a protein sequence $\mathbf{x}$ with maximum length $m$ (shorter sequences are padded to reach length $m$) partitioned into $T$ groups of $k$-mers, we use two types of RNNs (specifically, LSTMs here) in hierarchy for modeling within and across $k$-mers.  We first use an embedding layer to represent the $i$\tss{th} residue in $t$\tss{th} $k$-mer as a vector $\mathbf{x}_{it}$.  And we use a shared LSTM for all $k$-mers for the latent representation of the residue: $\mathbf{h}_{it} = \mathrm{LSTM}(\mathbf{x}_{it})$ ($t=1,\ldots,T$).  We then summarize each $k$-mer as $\mathbf{k}_t$ with an intra-$k$-mer attention mechanism:   
\begin{equation}
    \begin{split}
        u_{it} &= \mathbf{v_1}\mathrm{tanh}(\Theta_1\mathbf{h}_{it}+\mathbf{b_1}) \,\,\, \forall \, \mathrm{i,t}\\
        u'_{it} &= \frac{ \mathrm{exp}(u_{it})}{\sum_{i'} \mathrm{exp}(u_{i't})} \,\,\, \forall \, \mathrm{i,t} \\
        \mathbf{k}_t &= \sum_{i} u'_{it} \mathbf{h}_{it} \,\,\, \forall \, \mathrm{t}
    \end{split}
\end{equation}
Then we use another LSTM for for $\mathbf{k}_t$ and reach $\mathbf{h}_t=\mathrm{LSTM}(\mathbf{k_t})$ ($t=1,\ldots,T$). 

The final representation for a protein sequence is the collection of $\mathbf{h}_t$.%

\subsubsection{Joint attention over protein-compound atomic pairs for interpretability.}
Once the representation of protein sequences ($\mathbf{h}_t$  where $t=1,\ldots,T$ is the index of protein $k$-mer) and that of compound sequences or graphs ($\mathbf{y}_j$  where $j=1,\ldots,n$ is the index of compound atom) are defined, they are processed with a joint $k$-mer--atom attention mechanism to interpret any downstream prediction: 
\begin{equation}
    \begin{split}
        N_{tj} &= \mathrm{tanh}( \mathbf{h}_t \Theta_2 \mathbf{y}_{j}) \,\,\, \forall \, \mathrm{t,j}\\
        \mathcal{W}_{tj}' &= \frac{\mathrm{exp}(\mathbf{N}_{tj})}{\sum_{t',j'}\mathrm{exp}(\mathbf{N}_{t'j'})} \,\,\, \forall \, \mathrm{t,j}
    \end{split}
\end{equation}
With $\mathcal{W}_{tj}'$, the joint attention between the $t$\tss{th} k-mer and the ${j}$\tss{th} atom, we can combine it with the intra-$k$-mer attention over each residue $i$ in the $t$\tss{th} k-mer and reach $\mathcal{W}_{ij}$, the joint attention between the $i$\tss{th} protein residue and the $j$\tss{th} compound atom: 
\begin{equation}
    \mathcal{W}_{ij} = u_{it}' \mathcal{W}_{tj}' \,\,\, \forall \, \mathrm{i,j}
\end{equation}
This joint attention mechanism is an extension of our previous work \cite{karimi2018deepaffinity} where a protein sequence was represented as a single, ``flat'' RNN rather than multiple,  hierarchical RNNs.

\subsection{DeepRelations}

\subsubsection{Overall architecture.} 
\begin{figure}[!htb]
    \centering
    \includegraphics[width=\textwidth]{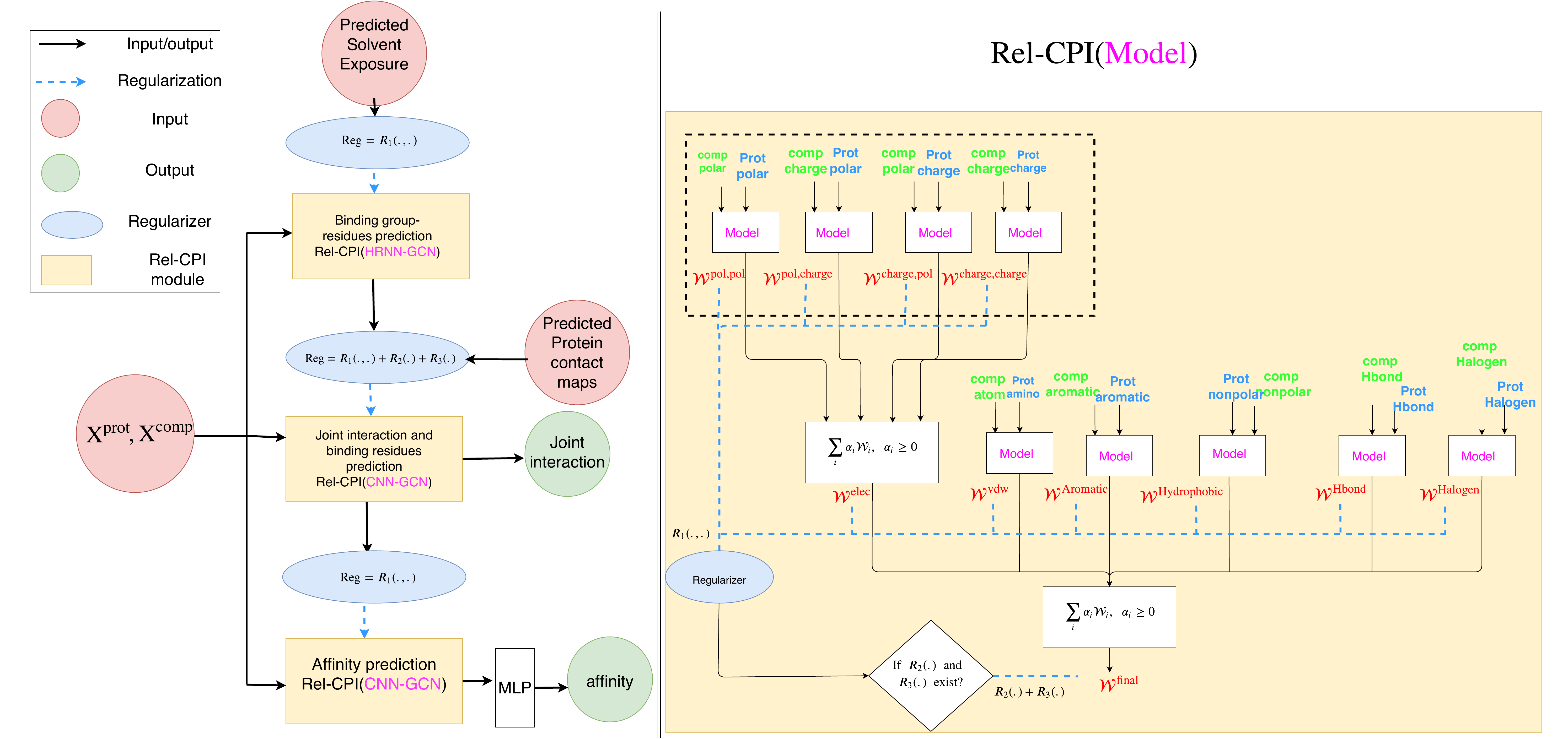}
    \caption{Schematic illustration of DeepRelations, an intrinsically explainable neural network architecture for predicting compound-protein interactions.}
    \label{fig:deeprelation}
\end{figure}

We have developed an end-to-end ``by-design" interpretable architecture named DeepRelations for joint prediction and  interpretation of compound-protein affinity. The overall architecture is shown in Figure~\ref{fig:deeprelation}. 

There are three relational modules (Rel-CPI) corresponding to three stages. Their attentions are trained to progressively focus on putative binding $k$-mers, residues, and pairs; and earlier-stage attentions guide those in the next stage through regularization. In each Rel-CPI module, there are six types of atomic ``relations" or interactions (including electrostaics as the non-negative linear combination of four sub-types). And each (sub)type of relation is modelled by aforementioned neural network pairs with joint attentions. For instance, the first Rel-CPI uses HRNN-GCN (HRNN for protein sequences and GCN for compound graphs) and the next two use CNN-GCN (dilated causal CNN for proteins and GCN for compounds).  And the non-negative linear combination of six individual relations' attention matrices %
$\mathcal{W}_{i}$ (where $i$ corresponds to the six relation types) 
gives the overall joint attention matrix $\mathcal{W}_\mathrm{final}$ (or $\mathcal{W}$ in short) in each module.

\subsubsection{Physics-inspired relational modules}

The relational modules are inspired by physics.  Specifically, atomic ``relations" or interactions constitute the physical bases and explanations of compound-protein interaction affinities and are often explicitly modelled in force fields. We have considered the following six types relations with attentions paid on and additional input data defined for.

\begin{itemize}
    \item \textit{Electrostatic interactions}: A non-negative linear combination of 
four subtypes of compound-protein interactions through attentions: 1) charge-charge, 2) charge-dipole, 3) dipole-charge, and 4) dipole-dipole interactions.  The input feature for the charge of a protein residue or a compound atom is the CHARMM27 parameter and the atomic formal charge, respectively.  That for the dipole of a protein residue or a compound atom is the residue being polar/nonpolar or the Gasteiger atomic partial charge.  
    \item \textit{Hydrogen bond}: Non-covalent interaction between an electronegative atom as a hydrogen ``acceptor'' and a hydrogen atom that is covalently bonded to an electronegative atom called a hydrogen ``donor''. Therefore,  if a protein residue or compound atom could provide a hydrogen acceptor/donor, its hydrogen-bond feature is -1/+1; otherwise the feature value is 0. A protein residue is allowed to be both hydrogen-bond donor and acceptor.  
    \item \textit{Halogen bond}: A halogen bond is very similar to hydrogen bond except that a halogen (rather than hydrogen) atom (often found in drug compounds) is involved in such interactions.  
    If a protein residue or a compound atom has/is a 
    halogen atom such as iodine, bromine, chlorine and fluorine, its halogen-bond feature is assigned +4, +3, +2 and +1, respectively, for decreasing halogen-bonding strength. If it can be a halogen acceptor, the feature is -1. If it can be neither, the feature is simply set at 0.
    \item \textit{Hydrophobic interactions}: The interactions between hydrophobic protein residues and compound atoms contribute significantly to the binding energy between them. 
    If a protein residue is hydrophobic,  it is represented as 1 and otherwise as 0. Moreover, the non-polar atoms are represented as 1 and the polar one as -1.
    \item \textit{Aromatic interactions}: Aromatic rings in tryptophans, phenylalanines, and tyrosines participate in ``stacking'' interactions with aromatic moieties of a compound ($\pi$-$\pi$ stacking). Therefore, if a protein residue has an aromatic ring, its aromatic feature is set at 1 and otherwise at 0. Similarly, if a compound atom is part of a ring, the feature is set at 1 and otherwise at 0. 
    \item \textit{VdW interactions}: Van der Waals are weaker interactions compared to others.  But the large amount of these interactions   contribute significantly to the overall binding energy between a protein and a compound. We consider the amino-acid type and the atom element as their features and use an embedding layer to derive their continuous representations.
\end{itemize}

 For each (sub)type of atomic relations, corresponding protein and compound features are fed into basis neural network models such as HRNN for protein sequences and GNN for compound graphs.   
 All features are made available to baseline methods (DeepAffinity+ variants) as well for fair comparison.    

\subsubsection{Physical constraints as regularization.}
The joint attention matrices $\mathcal{W}$ in each Rel-CPI module, for individual relations or overall, are regularized with the following two types of physical constraints.  

\paragraph{Focusing regularization}
In the first regularization, a constraint input is given as a matrix $\mathcal{T}\in [0,1]^{m\times n}$ to penalize the attention matrices $\mathcal{W}_i$ for all the 10 (sub)types of relations if they focus on the undesired regions of proteins.
In addition, an L1 sparsity regularization is on the attention matrices $\mathcal{W}_i$ for all relations to promote interpretability  as a small portion of protein residues interact with compounds. Therefore, this ``focusing" penalty can be formalized as:
\begin{equation}
    R_1(\mathcal{W}) = \lambda_\mathrm{relation}\sum_{i=1}^{10}|| (\bm{1}-\mathcal{T})\odot \mathcal{W}_i||_2 +  \lambda_\mathrm{L1} \sum_{i=1}^{10} ||\mathcal{W}_i||_1,
\end{equation}
where the $\mathcal{T}$ term, a parameter, can be considered as soft thresholding and its penalty only incurs when $\mathcal{T}_{ij}=0$.  

The first regularization is used for all three Rel-CPI modules or stages with increasingly focusing $\mathcal{T}$.  Let $\mathcal{T}^{[k]}$ be the constraint matrix  and $\mathcal{W}^{[k]}$ the learned attention matrix for a given relation in the $k$\tss{th} stage. In the first stage, $\mathcal{T}^{[1]}_{ij}$ is one only for any residue $i$ predicted to be solvent-exposed in order to focus on surfaces.  In the second stage,  $\mathcal{T}^{[2]}_{ij}=\max_{j^\prime} \mathcal{W}^{[1]}_{ij^\prime}$ to focus on putative binding residues hierarchically learned for $k$-mers and residues in stage 1.  In the last stage,  $\mathcal{T}^{[3]}_{ij}=\mathcal{W}^{[2]}_{ij}$ focuses on putative contacts between protein residues and compound atoms. The focusing regularization is enforced on attentions for every relation (sub)type in the current implementation and can be done only for given (sub)types in future.

\paragraph{Structure-aware sparsity regularization over protein contact maps}
We further develop a structure aware sparsity constraints based on known or RaptorX-predicted contact maps of the unbound protein. As sequentially distant residues might be close in 3D and form binding sites for compounds, we define overlapping groups of residues where each group consists of a residue and its spatially close neighboring residues.  Just in the second stage, we introduce Group Lasso for spatial groups and the Fused Sparse Group Lasso (FSGL) for sequential groups on the overall, joint attention matrix $\mathcal{W}$: 
\begin{equation}
    R_2(\mathcal{W}) =  \lambda_\mathrm{group} ||\mathcal{W}||_\mathrm{group} + \lambda_\mathrm{fused} ||\mathcal{W}||_\mathrm{fused} + \lambda_\mathrm{L1-overall} ||\mathcal{W}||_\mathrm{1}.
\end{equation}
The group Lasso penalty will encourage a structured group-level sparsity so that few clusters of spatially close residues share similar attentions within individual clusters. The fused sparsity will encourage local smoothness of the attention matrix so that sequentially close residues share similar attentions with compound atoms.  The L1 term maintains the sparsity of the overall attention matrix $\mathcal{W}$, since the L1 sparsity of attention matrices $\mathcal{W}_i$ for individual relations do not guarantee that their linear combination remains sparse. 

\subsubsection{Supervised attention.} 

It has been shown in visual question answering that attention mechanisms in deep learning can differ from human attentions\cite{das2017human}. As will be revealed in our results, they do not necessarily focus on actual atomic interactions (relations) in compound-protein interactions either.  We have thus curated a relational subset of our compound-protein pairs with affinities, for which known ground-truth atomic contacts or relations are available.  We summarize actual contacts of a pair in a matrix  $\mathcal{\overline{W}}^\mathrm{true}$ of length $\overline{m}\times\overline{n}$ ($\overline{m}$ and $\overline{n}$ are actual numbers of protein residues and compound atoms, respectively, for a given pair), which is a binary pairwise interaction matrix normalized by the total number of nonzero entries.  
We have accordingly introduced a third regularization term to  supervise $\mathcal{\overline{W}}$,  the non-padded submatrix of attention matrix $\mathcal{W}$, in the second stage: 
\begin{equation}
    R_3(\mathcal{W}) = \lambda_\mathrm{bind} \,\,c \,\,\,||\mathcal{\overline{W}} - \mathcal{\overline{W}}^\mathrm{true}||^2_{F}, 
\end{equation}
where $c$ is a normalization constant across batches.  
Suppose that, in any batch, a given pair's actual interaction matrix is of size $m_1 \times n_1$ and the smallest such size across all batches is $m_\mathrm{min} \times n_\mathrm{min}$.  
Then $c = \frac{m_\mathrm{min} \times n_\mathrm{min}}{m_1 \times n_1}$ for this pair.

\subsubsection{Training strategy for hierarchical multi-objectives}
Accuracy and interpretability are the two objectives we pursue at the same time. In our case, the two objectives are hierarchical: compound-protein affinity originates from atomic-level interactions (or ``relations'') and better interpretation in the latter potentially contributes to better prediction of the former.   

Challenges remain in solving the hierarchical multi-objective optimization problem.  First, optimizing for both objectives simultaneously (for instance, through weighted sum of them) does not respect that the two objectives do no perfectly align with each other and are of different sensitivities to model parameters.  Second, ground-truth data for interpretability of affinity prediction, i.e., compound-protein contacts, is rare.  In fact, merely 7.5\% of our compound-protein pairs labeled with $K_d$ affinities are with contact data.  

To overcome the aforementioned challenges, we consider the problem as multi-label machine learning facing missing labels.  And we design hierarchical training strategies to solve the corresponding hierarchical multi-objective optimization problem.  The whole DeepRelations model, including the three Rel-CPI modules, are trained end-to-end \cite{wang2016studying}.  In the first stage, we ``pre-trained" DeepRelations to minimize mean squared error (MSE) of p$K_d$ regression alone, with physical constraints turned on; in other words, attentions were regularized (through $R_1(\cdot)$ and $R_2(\cdot)$) but not supervised in this stage.  We tuned combinations of all hyperparameters except $\lambda_\mathrm{bind}$ in the discrete set of \{$10^{-5}, 10^{-4},\ldots,,10^{-1}$\}, with 400 epochs at the learning rate of 0.001.  Over the validation set, we recorded the lowest RMSE for affinity prediction and chose the hyperparameter combination with the highest AUPRC for contact prediction subjective to that the corresponding affinity RMSE (root mean square error) does not deteriorate from the lowest by more than 10\%.  

In the second stage, with the optimal values of all hyperparameters but $\lambda_\mathrm{bind}$ fixed, we loaded the corresponding optimized model in the first stage and ``fine-tuned" the model to minimize MSE additionally regularized by supervised attentions (through $R_1(\cdot)$, $R_2(\cdot)$, and $R_3(\cdot)$).  As only 7.5\% training examples are with known contacts, we used the their average and ignored the other examples for $R_3(\cdot)$ in each batch.  We used a slower learning rate (0.0001) and less training epochs (200) in the fine-tuning stage; and we tuned $\lambda_\mathrm{bind}$ in the set of \{$10^{-1}, 10^{-4},\ldots,,10^{-5}$\} following the same strategy as in the pre-training stage.  

In the end, we chose $\lambda_\mathrm{relation}=10^{-3}$, $\lambda_\mathrm{L1}=10^{-5}$, $\lambda_\mathrm{group}=10^{-2}$, $\lambda_\mathrm{fused}=10^{-3}$, $\lambda_\mathrm{L1-overall}=10^{-5}$ and $\lambda_\mathrm{bind}=10^1$ for DeepRelations.

We did similarly for hyper-parameter tuning while constraining (and supervising) attentions to make DeepAffinity+ variants.  
For HRNN-GCN\_cstr (modeling protein sequences with HRNN and compound graph with GCN, regularized by physical constraints in $R_2(\cdot)$), we chose  $\lambda_\mathrm{group}=10^{-5}$, $\lambda_\mathrm{fused}=10^{-4}$, and $\lambda_\mathrm{L1-overall}=10^{-5}$; and for its supervised version HRNN-GCN\_cstr\_sup, the additional $\lambda_\mathrm{bind}=10^4$.  For HRNN-GIN\_cstr (modeling protein sequences with HRNN and compound graph with GIN, regularized by physical constraints in $R_2(\cdot)$), we chose  $\lambda_\mathrm{group}=10^{-3}$,$\lambda_\mathrm{fused}=10^{-4}$, and $\lambda_\mathrm{L1-overall}=10^{-4}$; and for its supervised version HRNN-GIN\_cstr\_sup, the additional $\lambda_\mathrm{bind}=10^4$.  $R_1(\cdot)$ was for attentions on individual relations in DeepRelations and not applicable for DeepAffinity+ variants, although a surface-focusing regularization on overall attentions could be introduced.

\section{Results}

\subsection{Attentions alone are inadequate for interpreting compound-protein affinity prediction}

Our first task is to systematically assess the adequacy of attention mechanisms for interpreting model-predicted compound-protein affinities.  To that end, we adopt various data representations and corresponding state-of-the-art neural network architectures in our framework of DeepAffinity.   To model proteins, we have adopted RNN using protein SPS \cite{karimi2018deepaffinity} as input data as well as CNN and newly developed HRNN using protein amino-acid sequences.  To model compounds, we have adopted RNN using SMILES as input data as well as GCN and GIN using compound graphs with node features and edge adjacency  \cite{li2019deepchemstable}.  In the end, we have tested six DeepAffinity variants for protein-compound pairs, including RNN-RNN, RNN-GCN, CNN-GCN, HRNN-RNN, HRNN-GCN, and HRNN-GIN. The first two (RNN-RNN and RNN-GCN), where protein SPS sequences are modeled by RNN and compound SMILES or graphs are modeled by RNN or GCN, are essentially our previous models \cite{karimi2018deepaffinity} except that no unsupervised pretraining is used in this study. Whereas these two models' attentions on proteins are at the secondary structure levels (thus not assessed for interpretability here), the rest have joint attentions at the level of pairs of protein residues and compound atoms.  

\begin{figure}[!htb]
    \centering
    \includegraphics[width=.9\textwidth]{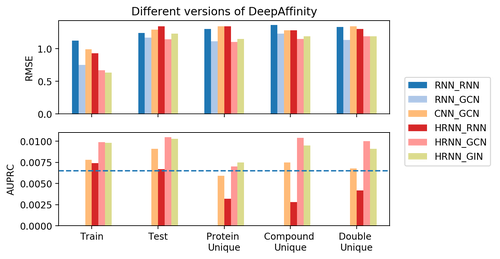}
    \caption{Comparing accuracy and interpretability among various versions of DeepAffinity with (unsupervised) joint attention mechanisms. Separated by underscores in legends are neural network models for proteins and compounds respectively.}
    \label{fig:comp}
\end{figure}

The accuracy of affinity prediction, measured by RMSE (root mean squared error) in p$K_d$, is summarized for the DeepAffinity variants in the top panel of Figure~\ref{fig:comp}.  Overall, all variants have shown p$K_d$ error between 1.1 and 1.3, a level competitively comparable even to the state-of-the-art affinity predictor using compound-protein co-crystal structures \cite{jimenez2018k}.  These models have robust accuracy profiles across the default, compound-unique, protein-unique, and double-unique test sets, suggesting their generalizability beyond training compounds or proteins.  Modeling compound SMILES with RNN seems to have slightly worse performance compared to modeling compound graphs with GCN or GIN, although less features are used for SMILES strings compared to node features for compound graphs.

The interpretability of affinity prediction is assessed against ground truth of contacts, as in the bottom panel of Figure~\ref{fig:comp}.  Specifically, we use joint attention scores to classify all possible residue-atom pairs into contacts or non-contacts.  As contacts only represent a tiny portion (0.0061$\pm$0.0023 in our dataset) of all possible pairs, we use the area under the precision-recall curve (AUPRC), instead of the area under the receiver operating characteristic curve (AUROC), to assess such binary classification. Here AUPRC is averaged over all pairs involved in the corresponding set.  Interestingly, compared to chance (AUPRC=0.0061), modeling protein amino-acid sequences through CNN or modeling compound SMILES through RNN had comparable or even worse contact prediction (or interpretability here).  Modeling protein amino-acid sequences through hierarchical RNN and compound graphs as GNN (GCN or GIN here) would improve the AUPRC by around 50\% for all test sets except the protein-unique one.  However, even with 50\% relative improvement, the absolute accuracy level of contact prediction, or the absolute level of model interpretability, remains low (AUPRC around 0.01).  Moreover, unlike the case of affinity accuracy, the interpretability results have shown some sensitivity to training data, especially when the test proteins are not contained in the training set.  

From the results above, we conclude that attention mechanisms alone are inadequate for the interpretability of compound-protein affinity predictors, regardless of the choice of commonly used, generic neural network architectures.

\subsection{Regularizing attentions with physical constraints modestly improves interpretability.}

Our next task is to enhance the interpretability of compound-protein affinity prediction beyond the level achieved by attention mechanisms alone.  The first idea is to incorporate domain-specific physical constraints into model training.  The rationale is that, by bringing in the (predicted) structural contexts of proteins and protein-compound interactions, attentions can be guided in their sparsity patterns  accordingly for better interpretability. 

\begin{figure}[!htb]
    \centering
    \includegraphics[width=.9\textwidth]{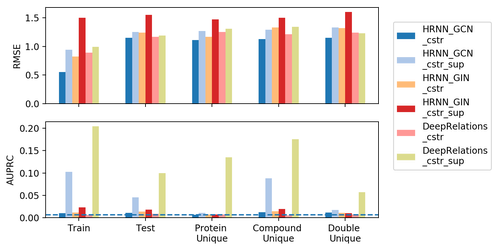}
    \caption{Comparing accuracy and interpretability among various versions of DeepAffinity+ (DeepAffinity with regularized and supervised attentions) and DeepRelations. ``cstr'' in legends indicates physical constraints imposed on attentions through regularization term $R_2(\cdot)$, whereas ``sup'' indicates supervised attentions through regularization term $R_3(\cdot)$.}
    \label{fig:comp_super}
\end{figure}

We start with the two best-performing DeepAffinity variants so far (HRNN-GCN and HRNN-GIN) where protein amino-acid sequences are modeled by hierarchical RNN and compound graphs by various GNNs (including GCN and GIN).  And we introduce 
structure-aware sparsity regularization $R_2(\cdot)$ to the two models to make ``DeepAffinity+" variants.  The resulting  HRNN-GCN\_cstr and HRNN-GIN\_cstr models with physical constraints are assessed in Figure~\ref{fig:comp_super}. 
Compared to the the non-regularized counterparts in Figure~\ref{fig:comp}, both models achieved similar accuracy levels across various test sets for affinity prediction, but their interpretability improved.  Specifically, HRNN-GCN after constraints, compared to that before constraints, had AUPRC improvement of 5.7\%, 2.9\%, 19.2\%, and 20.0\% for default test, protein-unique, compound-unique, and double-unique sets, respectively.  
However, the interpretability improvements from physical constraints were modest especially when the absolute level of AUPRC remained around 0.01. %
These results suggest that incorporating physical constraints to structurally regularize the sparsity of attentions is useful for improving interpretability but may not be enough.

\subsection{Supervising attentions significantly improves interpretability.} 
As regularizing attentions with physical constraints was not enough to enhance interpretability, our next idea is to additionally supervise attentions with ground-truth contact data available to some but not all training examples.  Again we introduce ``DeepAffinity+" models starting with HRNN-GCN and HRNN-GIN, by both regularizing and supervising attentions (using $R_2(\cdot)$ and $R_3(\cdot)$).   

The performances of resulting HRNN-GCN\_cstr\_sup and HRNN-GIN\_cstr\_sup models are shown in Figure~\ref{fig:comp_super}.  Importantly, HRNN-GCN\_cstr\_sup (light blue) significantly improves interpretability of affinity prediction without the sacrifice of accuracy.  The average AUPRC improved to 0.0455, 0.0106, 0.0883, and 0.0175 for the default test, protein-unique, compound-unique, and double-unique test sets, representing a relative improvement of 309\% (645\%), 92\% (73\%), 600\% (1347\%), and 46\% (186\%), respectively, compared to the constrained counterparts (chance). Interestingly, supervising attentions in HRNN-GIN did not see as significant improvement in interpretability.  

\subsection{Building explainability into DeepRelations architecture further drastically improves interpretability.}
Toward better interpretability, besides regularizing and supervising attentions, we have further developed an explainable, deep relational neural network named DeepRelations.  Here atomic ``relations'' constituting physical bases and explanations of compound-protein affinities are explicitly modeled in the architecture with multi-stage gradual ``zoom-in" to focus attentions.  In other words, the model architecture itself is intrinsically explainable by design.  

The superior performances of the resulting DeepRelations (with both regularized and supervised attentions) are shown in Figure~\ref{fig:comp_super} (yellow-green ``DeepRelations\_cstr\_sup'').  With equally competitive accuracy in affinity prediction as all previous models, DeepRelations achieved drastic improvements in interpretability. Strikingly, the average AUPRC further improved to 0.0996, 0.1350, 0.1754, and 0.0571 for the default test, protein-unique, compound-unique, and double-unique test sets, representing a relative improvement of 121\% (1532\%), 1173\% (2113\%), 98\% (2775\%), and 226\% (836\%), respectively, compared to the previous best DeepAffinity+ variant (chance).

\begin{figure}[!htb]
    \centering
    \includegraphics[width=.9\textwidth]{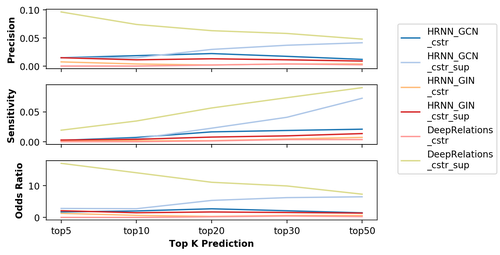}
    \caption{Comparing precision, sensitivity, and odds ratio (enrichment) of affinity-interpreting contacts predicted by various versions of DeepAffinity+ and DeepRelations.}
    \label{fig:comp_super_pso}
\end{figure}

We further assessed contact prediction (or interpretability) of DeepAffinity+ variants and DeepRelations using the precision, sensitivity, and odds ratio (or enrichment factor) of their top $K$ predictions (where $K$ ranged from 5 to 50).  Figure~\ref{fig:comp_super_pso} shows that DeepRelations drastically outperforms other methods in all assessment measures considered.  The precision and sensitivity levels may not appear impressive, largely due to the very strict definition of ``true contacts" in our study, as will be revealed in a case study.  Note that all atomic-level contact predictions were made with the inputs of protein sequences and compound graphs alone.  

For fair comparison, all DeepAffinity+ variants and DeepRelations were using the same set of features.  A negative control experiment in the subsequent ablation study further validated this.  Therefore, the architecture of DeepRelations, being intrinsically explainable, is the major contributor to its superior interpretability.  From the machine learning perspective, DeepAffinity+ variants have various molecular features  lumped into general-purpose neural networks, which makes it very hard to learn governing physics laws from the molecular affinity data.  Instead, DeepRelations directly builds the physics laws into its model architecture and carefully structure various features into corresponding atomic relations and eventually the overall binding affinity. 

\subsection{Ablation study for DeepRelations}

To disentangle various components of DeepRelations and understand their relative contributions to DeepRelations' superior interpretability, we removed components from DeepRelations and made  ``DeepRelations-" variants.  Besides regularized and supervised attentions, we believe that the main contributions in the architecture itself are (1) the multi-stage ``zoom-in" mechanisms that progressively focus attentions from surface, binding $k$-mers, binding residues to binding residue-atom pairs; and (2) the explicit modeling of atomic relations that can explain the structure feature-affinity mappings consistently with physics principles.  We thus made three DeepRelations- variants: DeepRelations without multi-stage focusing, without explicit atomic relations, or without both.  

\begin{figure}[!htb]
    \centering
    \includegraphics[width=.9\textwidth]{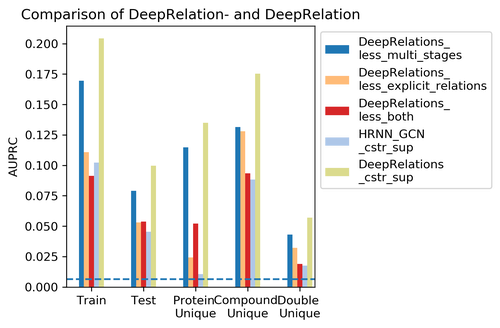}
    \caption{Comparing interpretability between DeepRelations and DeepRelations- (DeepRelations without multi-stage focusing, explicitly-modeled relations, or both).}
    \label{fig:ablation_study}
\end{figure}

We compared the three intermediate ``DeepRelations-" versions with the best DeepAffinity+ (regularized and supervised HRNN-GCN) and DeepRelations in Figure~\ref{fig:ablation_study}.  Consistent with our conjecture, we found that, the explicit modeling of atomic relations was the main reason for DeepRelations' superior interpretability, as the removal of this component alone reduced the average AUPRC down to a similar level of the best DeepAffinity+ (except for the protein-unique case). Removing both components essentially reproduced the best DeepAffinity+ (again, except that it still outperforms the latter in the protein-unique case), which served well as a ``negative control" case here.

\subsection{Case Study}

Now that we have established how drastically DeepRelations improves the interpretability of compound-protein affinity prediction and explained why it achieves so by design, we went on to examine the pattern in which DeepRelations contact prediction outperforms the best DeepAffinity variant HRNN-GCN (and leaves room for further improvement). We thus randomly chose a compound-protein test pair with known contacts for case study: carbonic anhydrase II inhibitor with its compound AL1 (PDB ID:1BNN).  

\begin{figure}[!htb]
    \centering
    \includegraphics[width=.32\textwidth]{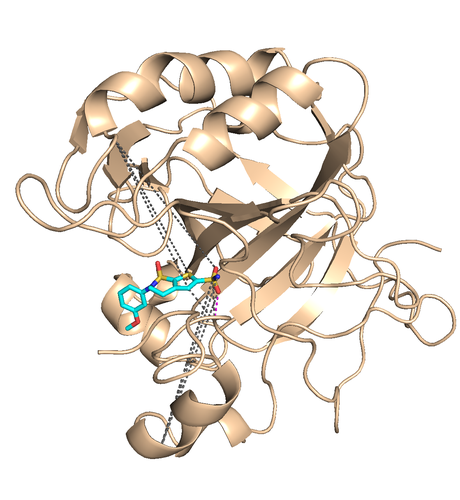}\hfill
    \includegraphics[width=.32\textwidth]{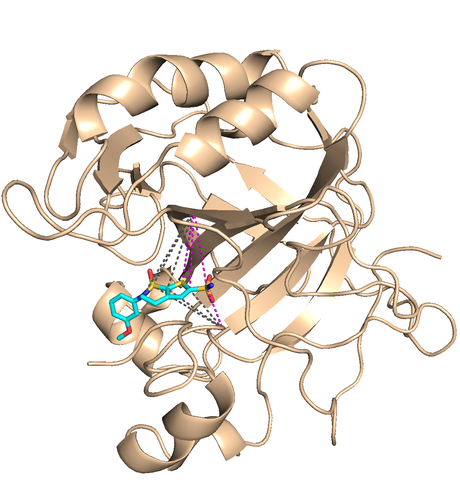}\hfill
     \includegraphics[width=.32\textwidth]{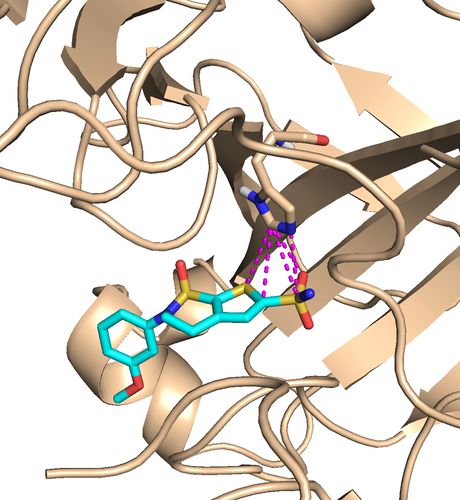}\hfill
    \caption{Structural visualization of top-10 intermolecular contacts predicted by HRNN-GCN (left) and DeepRelations (middle and right) for a test case. The protein (carbonic anhydrase II inhibitor) is shown in wheat cartoons and compound (AL1) in cyan sticks.  Dashed lines represent the top-10 predicted relations (interactions between protein residues and compound atoms) in either case. Their magenta and gray colors indicate correct and incorrect relations/contacts, respectively, according to ground truth strictly defined as in PDBsum.  In the right panel, we examined the dominant electrostaic attention matrix alone and found four true positives, all formed with Histidine 94 (wheat sticks), in the  top-10 predicted electrostatic contacts.  
    }
    \label{fig:case_study}
\end{figure}

As shown in Figure~\ref{fig:case_study}, DeepRelations (middle) not only made more correct contact predictions than HRNN-GCN (left) but also showed much improved contact \textit{pattern}. In particular, HRNN-GCN could focus attention on residue-atom pairs that are actually as far as above 20\AA\ away, and the attended residues could be dispersed at two sides of a protein.  In contrast, DeepRelations predictions were correctly focused in the binding site of the protein and many of its ``incorrect" predictions may correspond to residue-atom pairs within 10\AA\ or less, which could be partially attributed to the physical constraints introduced as regularization.

To further examine the possible benefit of explicitly modeled atomic relations, we examined the overall attention matrix and found that the most contributions originate from electrostatic relations.  We therefore examined the top-10 predicted electrostatic contacts according to the electrostatic attention matrix alone and found four true electrostatic interactions associated with the same protein residue (Hisidine 94). 

\begin{figure}[!htb]
    \centering
    \includegraphics[width=.9\textwidth]{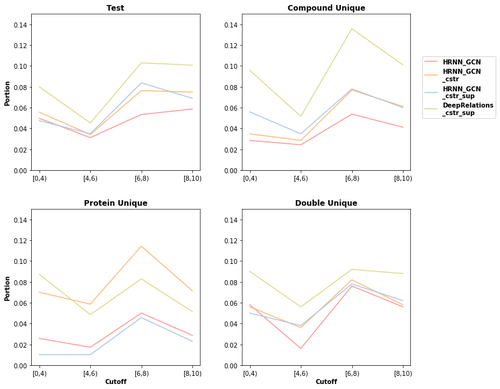}
    \caption{Distributions of top-50 contacts, predicted by DeepAffinity, DeepAffinity+, and DeepRelations, in various distance ranges (unit: \AA).}
    \label{fig:top50_all_test}
\end{figure}

We extended the analysis of the patterns of predicted contacts over all test cases.  Considering that the true contacts are defined rather strictly, we assess distance-distributions of residue-atom pairs predicted by HRNN-GCN, HRNN-GCN with regularized attention, HRNN-GCN with regularized and supervised attention, and DeepRelations (also with regularized and supervised attention).  As seen in Figure~\ref{fig:top50_all_test}, DeepRelations outperforms competitors in all distance ranges over all test sets (except the 4\AA$\sim$10\AA\ range for the seemingly most challenging protein-unique case).  Impressively, among top-50 contacts predicted by DeepRelations, around 33\%, 39\%, 27\%, and 33\% 
were actually within 10\AA in the default test, compound-unique, protein-unique, and double-unique test sets, respectively.   

\section{Conclusions}

Toward accurate and interpretable machine learning of compound-protein affinity, we have curated an affinity-labeled dataset with partially annotated contact details, assessed the adequacy of current attention-based deep learning models for both accuracy and interpretability, and developed novel machine-learning models and training strategies to drastically enhance interpretability without sacrificing accuracy.  This is the first study with dedicated model development and systematic model assessment for interpretability in affinity prediction. 

Our study has found that commonly-used attention mechanisms alone, although better than chance in most cases, are not satisfying in interpretability: the most attended contacts in affinity prediction do not reveal true contacts underlying affinities at a useful level.  We have tackled the challenge with three innovative, methodological advances.  First, we introduce domain-specific physical constraints to regularize attentions (or guide their sparsity patterns), in which structural contexts such as sequence-predicted protein surfaces and protein contact maps are utilized.  Second, we exploit partially available ground-truth contacts to supervise attentions.  Lastly, we build intrinsically explainable model architecture where various atomic relations, reflecting physics laws, are explicitly modeled and aggregated for affinity prediction.  Joint attentions are embedded over residue-atom pairs for individual and overall relations.  And a multi-stage hierarchy, trained end-to-end, progressively focuses attentions on protein surfaces, binding $k$-mers and residues, and residue-atom contact pairs.  

Empirical results demonstrate the superiority of DeepRelations in interpretability without sacrificing accuracy.  Compared to the best DeepAffinity variant with joint attention (HRNN-GCN), the AUPRC for contact prediction was boosted to  9.48, 16.86, 19.28, and 5.71-fold for the default test, compound-unique, protein-unique, and double-unique cases. Importantly, the interpretability of DeepRelations proves robust and generalizable, as the margins of improvement were even higher when compounds or/and proteins are not present in the training set. Ablation studies demonstrate that the explainable relational network architecture was the major contributor to such performances.  Case studies suggest that DeepRelations predict not only more correct but also more well-patterned contacts.  And many ``incorrect'' predictions due to the strict definition of contacts were within reasonable ranges --- in fact, around one third of the top-50 predicted contacts correspond to residue-atom pairs within 10\AA.    

An additional benefit of DeepRelations is its broad applicability toward the vast chemical and proteomic spaces. It does not rely on 3D structures of compound-protein complexes or even protein monomers when such structures are often unavailable.  The only inputs needed are protein sequences and compound graphs.  Meanwhile, it adopts the latest technology to predict structural contexts for protein sequences (such as surfaces, secondary structures, and residue-contact maps) and incorporatea such structural contexts into affinity and contact predictions.  When structure data are available, DeepRelations can readily integrate such data by using actual rather than predicted structural contexts.  

Our study demonstrates that, it is much more effective to directly build explainability into machine learning model architectures (as DeepRelations models underlying atomic relations explicitly) than to infer explainability from general-purpose architectures (as DeepRelations variants learn attentions from data alone).  In other words, designing intrinsically interpretable machine learning models, although more difficult, can be much more desired than pursuing interpretability in a \textit{post hoc} manner.

\begin{acknowledgement}
This work was supported by the National Institutes of Health (R35GM124952 to Y.S.).  Part of the computing time was provided by the Texas A\&M High Performance Research Computing. 
\end{acknowledgement}

\bibliography{Ref}

\end{document}